\def\real{{\tt I\kern-.2em{R}}}
\def\nat{{\tt I\kern-.2em{N}}}
\def\eps{\epsilon}
\def\realp#1{{\tt I\kern-.2em{R}}^#1}
\def\natp#1{{\tt I\kern-.2em{N}}^#1}
\def\hyper#1{\ ^*\kern-.2em{#1}}
\def\monad#1{\mu (#1)}

\def\st#1{{\tt st}(#1)}
\def\hyperreal{{^*{\real}}}
\def\hyperrealp#1{{\tt ^*{I\kern-.2em{R}}}^#1} 

\def\hypernatp#1{{{^*{{\tt I\kern-.2em{N}}}}}^#1} 
\def\eskip{\hskip.25em\relax}

\def\Hyper#1{\hyper {\eskip #1}}
\def\leaderfill{\leaders\hbox to 1em{\hss.\hss}\hfill}
\def\srealp#1{{\rm I\kern-.2em{R}}^#1}

\def\power#1{{{\cal P}(#1)}}

\def\parm{\par\medskip}

\def\ref#1{$^{#1}$}

\def\m@th{\mathsurround=0pt}
\def\rightarrowfill{$\m@th \mathord- \mkern-6mu \cleaders\hbox{$\mkern-2mu 
\mathord- \mkern-2mu$}\hfil \mkern-6mu \mathord\rightarrow$}
\def\leftarrowfill{$\mathord\leftarrow
\mkern -6mu \m@th \mathord- \mkern-6mu \cleaders\hbox{$\mkern-2mu 
\mathord- \mkern-2mu$}\hfil $}
\def\noarrowfill{$\m@th \mathord- \mkern-6mu \cleaders\hbox{$\mkern-2mu 
\mathord- \mkern-2mu$}\hfil$}
\def\orgate{$\bigcirc \kern-.80em \lor$}
\def\andgate{$\bigcirc \kern-.80em \land$}
\def\inverter{$\bigcirc \kern-.80em \neg$}

\magnification=1200
                %IJMMS FORMATE  TYPOS CORRECTED 16 MAY 95
\tolerance 10000
\hoffset=0.25in  %IJMMS format
\baselineskip=14pt
%\hsize 7 true in  %IJMMS format
\hsize 6.0 true in
%\vsize 11.20 true in  %IJMMS format
\vsize 8.5 true in
%\nopagenumbers  %IJMMS format
{\quad}
%\vskip 29.73pt %IJMMS format
\line{\hfil \bf A Hypercontinuous Hypersmooth\hfil} 
\line{\hfil \raise4pt\hbox{\bf Scharzschild Line Element Transformation}\hfil}
\bigskip
\centerline{\bf Robert A. Herrmann}
%\line{\hfil {Mathematics Department} \hfil}
%\line{\hfil\raise5pt\hbox{United States Naval Academy} \hfil}
%\line{\hfil\raise12pt\hbox{572 Holloway Rd.}\hfil}
%\line{\hfil\raise12pt\hbox{Annapolis MD 21402-5002 USA}\hfil} 
\baselineskip=12pt\centerline{Mathematics Department}
\centerline{United States Naval Academy}
\centerline{572C Holloway Rd.}
\centerline{Annapolis, MD 21402-5002 USA}
%\vskip 0.90 true in  

\baselineskip=14pt
\vskip 24pt
\noindent ABSTRACT. 
In this paper, a new derivation for one of the
 black hole line 
elements 
is given since the basic derivation for this line element is 
 flawed mathematically. This derivation postulates a transformation procedure
that utilizes a transformation function that is modeled by an ideal 
nonstandard physical world transformation process that  yields a 
connection between an exterior Schwarzschild line element and distinctly different
interior line element. The transformation is an ideal transformation in that in 
the natural world the transformation is conceived of as occurring  at an unknown 
moment in the evolution of a gravitationally  collapsing spherical body with 
radius greater than but near to the Schwarzschild radius. An ideal 
transformation models this transformation in a manner independent of the 
objects standard radius. It yields predicted behavior based upon a 
Newtonian gravitational field prior to the transformation, predicted 
behavior after the transformation for a field internal to the Schwarzschild 
surface and predicted behavior with respect to field alteration processes
during the transformation.  \par 
\par\bigskip\smallskip
\noindent {\bf 1.\hskip 1.25em Introduction.} \par
In [1], the linear effect line element is derived and,  
in [2], a general line element $dS^2$ is derived using the Special Theory 
chronotopic interval and two infinitesimal transformations  
(i) $dR^s = (1-\alpha\beta)dR^s - \alpha dT^s$ and (ii) $dT^s = \beta dR^m + dT^m,$ where the $\alpha$ 
and $\beta$ are to be determined. From these determinations, the following 
general line  element is derived.  
$$dS^2 = \lambda(cdt^{m})^2 - (1/\lambda)(dR^{m})^2 - 
(R^{s})^2(\sin^2\theta^{s}(d\phi^{s})^2 +
(d\theta^{s})^2). \eqno (1.1)$$\par
 The Eddington-Finkelstein transformation is the least ad hoc and is 
more physically justified than others. But, in [5], 
the derivation and argument for using the this simple 
transformation 
(1) $dU^m = dt^m + f_M(R^m)dR^m$ to obtain a black hole line element is flawed. 
This flaw is caused by the usual ad hoc logical 
errors in  ``removing infinities.'' Equation 57.11 in [5, p. 157], 
specifically requires that $R^m> 2GM/c^2.$ However, in arguing for the use of 
the transformed Schwarzschild line element (1.1), Lawden assumes that it is 
possible for 
$R^m= 2GM/c^2.$ But the assumed real valued function defined by 
equation 57.11 is not 
defined for $R^m= 2GM/c^2.$  Hence, a new and  
rigorously correct procedure is necessary. This is accomplished by showing 
that (1) can be considered as a hypercontinuous and hypersmooth 
transformation associated with a new $P$-process that 
yields an alteration to the gravitational  field in the vicinity of the 
Schwarzschild surface  during the process of gravitational collapse.\par This 
{\it speculation} 
is modeled by the expression (1) which is conceived of as an alteration in the 
time measuring light-clock. Further, this alteration is conceptually the same 
as the ultrasmooth microeffects model for fractual behavior [4]. This 
transformation takes the Schwarzschild line element, which  applies only 
to the case where $R^m> 
2GM/c^2,$ and yields an NSP-world 
 black hole line element that only applies for the case 
where $R^m\leq 2GM/c^2.$ Like ultrasmooth microeffects, the nonstandard 
transformation process is considered as an ideal model of behavior that 
approximates the actual natural world process.  
Thus we have two district line elements connected by such a transformation and each 
applies to a specific $R^m$ domain.\par 
\noindent {\bf 2.\hskip 1.25em The Function $f_M(R^m)$. }\par
To establish that an internal function $f_M(R^m)$ exists with the 
appropriate properties proceed as follows: let $\cal I$ be the set of all 
nonsingleton intervals in $\power{\real}.$ Let $\cal F\subset \power {\real \times 
\real}$  be the set of all nonempty functional sets of ordered pairs. For each 
$I \in \cal I,$ let $C(I,\real) \subset \cal F$ be the set of all real valued 
continuous functions (end points included as necessary) defined on $I.$ For 
each $a > 0,$  $\exists f_a\in C((-\infty,0],\real),\ (-\infty,0]\in
\cal I,$  such that $\forall x \in (-\infty,0],\ f_a(x) = 1/(x-a).$  Further,
$\exists g_a\in C((0,2a],\real),\ (0,2a]\in
\cal I,$  such that $\forall x \in (0,2a],\ g_a(x) = 
-x^3/(2a^4)+7x^2/(4a^3)-x/a^2-1/a.$ Then 
$\exists h_a\in C((2a,+\infty),\real),\  (2a,+\infty) \in
\cal I,$  such that $\forall x \in (2a,+\infty),\ h_a(x) = 0.$ Finally, it 
follows that $\lim_{x \to 0^-} f_a(x) = \lim_{x \to 0^+} g_a(x),\ 
\lim_{x \to 2a^-} g_a(x) =\lim_{x \to 2a^+} h_a(x).$ Hence 
$$H_a(x) = \cases{f_a(x);&$x\in (-\infty,0]$\cr
                g_a(x);&$x\in (0,2a]$\cr
                h_a(x);&$x\in (2a,+\infty)$\cr}$$ 
is continuous for each $x \in \real$ and has the indicated properties.\par
Now $H^\prime_a(x)$ exists and is continuous for all $x \in \real$ and 
$$H^\prime_a(x) = \cases{f_a^\prime(x);&$x\in (-\infty,0]$\cr
                g_a^\prime(x);&$x\in (0,2a]$\cr
                h_a^\prime(x);&$x\in (2a,+\infty)$\cr}$$ \par
All of the above can be easily expressed in a first-order language 
and all the statements hold in our superstructure enlargement [4]. Let $0<\eps \in 
\monad 0.$ Then there exists an internal hypercontinuous hypersmooth $H_\eps \colon 
\hyperreal \to \hyperreal$ such that $\forall x \in 
\Hyper {(-\infty,0]},\ 
H_\eps (x)= 1/(x - \eps)$ and  $\forall x \in  \Hyper {(-\infty,0)}\cap    
\real,\ \st {H_\eps (x)} =\st {1/(x - \eps )} = 1/x;$  and for $x = 0,\ 
H_\eps(0)$ exists, although $\st {H_\eps(0)}$ does not exist as a real 
number. Further, 
$\forall x \in  {(2\eps ,+\infty )}\cap \real=(0,+\infty),   
\ \st {H_\eps(x)} =0.$  To obtain the hypercontinuous hypersmooth $f_M,$  
simply let $cf_M 
= H_\eps,\ x = \lambda,\ R^m \in \hyperreal.$ \par
\noindent {\bf 3.\hskip 1.25em Motivation for Function Selection. }\par
Recall that a function $f$ defined on interval $I$ is 
standardizable (to $F$) on $I$ if 
$\forall x \in I \cap \real,\  F(x) = \st {f(x)}\in \real.$ 
Now, consider the transformation (1) in the nonstandard form 
$dU^m = dt^m +f_M(R^m)dR^m$ where internal $f_M(R^m)$ is a function defined 
on $A \subset \hyperreal,$ and $\lambda=\lambda(R^m).$ There are infinitely 
many nonstandard functions that can be standardized to produce the line 
element 
$dS^2.$ In this line element, consider substituting  for the function $\lambda = 
\lambda (R^m,)$ the function $\hyper \lambda -\eps.$ The transformed 
line element then  
becomes, prior to standardizing the coefficient functions (i.e. restricting 
them the the natural world), 
$$T= (\hyper\lambda -\eps)c^2((dU^m)^2 -2f_MdU^mdR^m + f^2_M(dR^m)^2) -
(1/(\hyper\lambda -\eps))(dR^m)^2-$$
$$(R^{m})^2(\sin^2\theta^{m}(d\phi^{m})^2 +
(d\theta^{m})^2)=$$ 
$$(\hyper\lambda -\eps)c^2(dU^m)^2 -2(\hyper \lambda -\eps)c^2f_MdU^mdR^m  + 
$$ $$
\overbrace{((\hyper\lambda -\eps)c^2 f^2_M- 
(1/(\hyper\lambda-\eps)))dR^m}^b dR^m -$$
$$(R^{m})^2(\sin^2\theta^{m}(d\phi^{m})^2 +
(d\theta^{m})^2).\eqno (3.1)$$ \par
Following the  procedure outlined in [4], 
first consider the partition 
$\real= (-\infty, 0] \cup (0,2\eps] \cup (2\eps, +\infty),$  where
$\eps$ is a positive infinitesimal. Consider the required constraints. 
(2) As required, for specific real intervals, 
all coefficients of the terms of the transformed
line element are to be standardized and, hence, are standard functions. 
(3) Since any  line element 
transformation, prior to standardization, should retain its infinitesimal 
character with respect to an 
appropriate interval $I$, then for any 
infinitesimal $dR^m$ and for each value $R^m\in I$ terms such as  
$G(R^m)dR^m,$ where $G(R^m)$ is a coefficient function, 
must be of  infinitesimal value.\par
For the important constraint (3), 
Definition 4.1.1, and theorems 4.1.1, 4.1.2 in [3] 
imply that for a fixed
infinitesimal $dR^m$  in order to have expression $b$ infinitesimal as $R^m$ varies, the 
coefficient $h(R^m)= (\hyper\lambda -\eps)c^2 f^2_M- 
1/(\hyper\lambda-\eps)$ must be infinitesimal on a subset $A$
of an appropriate interval $I$ such 
that  $0 \in A.$ The simplest case would be to assume that $A = \Hyper (-
\infty,0].$ Let 
standard $r \in A \cap \real.$ Then it follows that $h(r) \subset \monad 
0.$ Thus $\st {h(r)} = 0.$ Indeed, let $x \in (\cup \{\monad{r}\mid r <0,\ r 
\in \real \})\cup (\monad 0 \cap A).$ Then $\st {h(x)} = 0.$ Since we are seeking a transformation 
process that is  hypercontinuous, at least on $\Hyper (-\infty, 0],$ this last 
statement suggests the simplest to consider would be that  
 on $\Hyper (-\infty,0],\  h = 0.$ Thus the basic constraint yields the 
basic requirement that on $\Hyper (-\infty, 0]$ the simplest function to choose 
is $cf_M(x) =1/(x - \eps).$ Since standardizing is required 
on $\Hyper (-
\infty, 0)\cap \real,$ we have for each $x \in \Hyper (-
\infty, 0)\cap \real,$ that $\st {cf_M(x)} = c\st {f_M(x)} =\st {1/(x - 
\eps)}= 1/x.$
This leads to the assumption that 
on 
$(-\infty, 0]$ the function $f_a(x) = 1/(x - a),\ a >0,$ 
should be considered. 
After *-transferring and prior to standardizing, this selection
would satisfy (3) for both of the coefficients in which
$f_M$ appears and for the interval $I=\Hyper (-\infty,0].$ 
The function $g_a$ is arbitrarily selected to satisfy the hypercontinuous 
and hypersmooth property and, obviously, $h_a$ is selected to preserve 
the original  line element  for the interval $(2\eps, +\infty).$ Finally, it 
is necessary that the resulting new coefficient functions, prior to 
standarizing, all satisfy (3) at least for a fixed $dR^m$ and a varying 
$R^m\in \Hyper (-\infty, 0]$ for the expression (1).   
It is not difficult to show 
that $\vert H_\eps (x) \vert \leq 2/\eps$ for all $ x \in \hyperreal.$ 
Consequently,  for 
$\eps = (dR^m)^{1/3}$ expression (1) is an infinitesimal for all $R^m \in 
\hyperreal.$ \par
Let $1-v^2/c^2=\lambda.$ For the collapse scenario $R^m =R_M.$ If
$2GM/(R^mc^2) < 1,$ substituting $2GM/R^m = v^2,$ into (1.1) 
yields  
the so-called Schwarzschild line element. With respect to the transformation, 
(A) if $R^m <2GM/c^2,$ then for $ 
\st {f_M(R^m)} = 1/(c\lambda), 
\ \lambda = 1 - 2GM/(R^mc^2);$ 
for (B) $R^m > 2GM/c^2;\ \st {f_M(R^m)} = 0,$ and for the case that 
(C) $R^m = 2GM/c^2,$ 
the 
function $f_M$ is defined and equal to a NSP-world value $f_M(R^m).$  
But, for case  (C), $\st {f_M(R^m)}$ does not exist as a real number. 
 Hence, (C) 
has no direct effect within the natural world when $R^m = 2GM/c^2,$ although 
the fact that 
$f_M(R^m)dR^m$ 
is an infinitesimal implies that $\st {f_M(R^m)dR^m} = 0.$ Using 
these NSP-world functions and (3.1), cases (A) and (C) yield  
$$dS_1^2 = \lambda(cdU^{m})^2 - 2cdU^m\,dR^m -$$ 
$$(R^{m})^2(\sin^2\theta^{m}(d\phi^{m})^2 +
(d\theta^{m})^2). \eqno (3.2)$$
But case (B), leads to (1.1). The two constraints are met by 
$f_M(R^m),$ and indeed the standardized (A) form for $f_M(R^m)$ is unique if (3) is to be 
satisfied for a specific interval.\par
Since this is an ideal approximating model,  
in order to apply this ideal model to the natural world, 
one most select an 
appropriate real $a$ for the real valued function $H_a.$ Finally, it is not 
assumed that the function $g_a$ is unique. In any solutions for line element
(3.2), the $dU^m$ [resp. $dR^m$] refers to the timing [resp. 
length] infinitesimal light-clock counts and does refer to 
universal time [resp . length] alterations. \par
\bigskip\medskip
\baselineskip=12pt\centerline{{\bf Reference}} \par\medskip\smallskip
\noindent 1. Herrmann, R.A. An operator equation and 
relativistic alterations in the time for radioactive decay, {\it Intern. J. Math. \& Math. Sci.}, 19(2)(1996), 397-402.
 \parm\smallskip   
\noindent 2. Herrmann, R.A. {\it Constructing Logically Consistent Special and General Theories of Relativity}, Math. Dept., U.S. 
Naval Academy, Annapolis, MD, 1993. \parm\smallskip
\noindent 3. Herrmann, R.A. {\it Some Applications of Nonstandard Analysis to Undergraduate Mathematics: 
Infinitesimal Modeling and Elementary Physics},
{Instructional} {Development} {Project}, 
 {Mathematics} {Department}, U. S. Naval Academy, Annapolis, MD, 21402-5002, 
1991.
\parm\smallskip
\noindent 4. Herrmann, R.A. Fractals and ultrasmooth microeffects, 
{\it J. Math. Phys.}, 30(April 1989),
805-808.\parm\smallskip                                  
\noindent 5. Lawden, D. F. {\it An Introduction to 
Tensor Calculus, Relativity and Cosmology}, John Wiley 
 \& Sons, New York, 1982.
\end